\documentclass[reprint,superscriptaddress,aps]{revtex4-1}
\usepackage{amsmath}
\usepackage{comment}
\usepackage{mathtools}
\usepackage{amssymb}
\usepackage{bm}
\usepackage{braket}
\usepackage{color}
\usepackage[varg]{txfonts}
\usepackage{varwidth}
\usepackage{dcolumn}
\usepackage[breaklinks,colorlinks=true,linkcolor=blue,urlcolor=cyan,citecolor=blue]{hyperref}
\usepackage{graphicx}

\newcommand{\PZO}{Pr$_2$Zr$_2$O$_7$}
\newcommand{\HTO}{Ho$_2$Ti$_2$O$_7$}

\begin{document}
\title{Crystal field magnetostriction of spin ice under ultrahigh magnetic fields}

\author{Nan Tang\textsuperscript{*}}
\email{nan.tang@uni-a.de}
\affiliation{Experimental Physics VI, Center for Electronic Correlations and Magnetism, University of Augsburg, Augsburg 86159, Germany}

\author{Masaki Gen\textsuperscript{*}}
\email{gen@issp.u-tokyo.ac.jp}
\affiliation{Institute for Solid State Physics, University of Tokyo, Kashiwa, Chiba 277-8581, Japan}

\thanks{\textsuperscript{*}These authors contributed equally to this work.}

\author{Martin Rotter}
\affiliation{McPhase Project, Dresden 01159, Germany}

\author{Huiyuan Man}
\affiliation{Geballe Laboratory for Advanced Materials, Stanford University, California 94305, USA}

\author{Kazuyuki Matsuhira}
\affiliation{Faculty of Engineering, Kyushu Institute of Technology, Kitakyushu, Fukuoka 804-8550, Japan}

\author{Akira Matsuo}
\affiliation{Institute for Solid State Physics, University of Tokyo, Kashiwa, Chiba 277-8581, Japan}

\author{Koichi Kindo}
\affiliation{Institute for Solid State Physics, University of Tokyo, Kashiwa, Chiba 277-8581, Japan}

\author{Akihiko Ikeda}
\affiliation{Institute for Solid State Physics, University of Tokyo, Kashiwa, Chiba 277-8581, Japan}
\affiliation{Department of Engineering Science, University of Electro-Communications, Chofu, Tokyo 182-8585, Japan}

\author{Yasuhiro H. Matsuda}
\affiliation{Institute for Solid State Physics, University of Tokyo, Kashiwa, Chiba 277-8581, Japan}

\author{Philipp Gegenwart}
\affiliation{Experimental Physics VI, Center for Electronic Correlations and Magnetism, University of Augsburg, Augsburg 86159, Germany}

\author{Satoru Nakatsuji}
\affiliation{Institute for Solid State Physics, University of Tokyo, Kashiwa, Chiba 277-8581, Japan}
\affiliation{Department of Physics, University of Tokyo, Tokyo 113-0033, Japan}
\affiliation{Institute for Quantum Matter and Department of Physics and Astronomy, Johns Hopkins University, Baltimore, Maryland 21218, USA}
\affiliation{Trans-scale Quantum Science Institute, University of Tokyo, Tokyo  113-0033, Japan}

\author{Yoshimitsu Kohama}
\affiliation{Institute for Solid State Physics, University of Tokyo, Kashiwa, Chiba 277-8581, Japan}

\begin{abstract}
We present a comprehensive study of the magnetoelastic properties of the Ising pyrochlore oxide \HTO{}, known as spin ice, by means of high-field magnetostriction measurements and numerical calculations.
When a magnetic field is applied along the crystallographic $\langle 111 \rangle$ axis, the longitudinal magnetostriction exhibits a broad maximum in the low-field regime around 30~T, followed by a dramatic lattice contraction due to crystal-field (CF) level crossing at $B_{\rm cf} \sim 65$~T.
The transverse magnetostriction exhibits a contrasting behavior, highlighting the anisotropic nature of the CF striction.
By applying a magnetic field at varying sweep rates, we identify distinct timescales of spin dynamics that are relevant to monopole formation and annihilation, as well as CF-phonon dynamics.
Our mean-field calculations, based on a point-charge model, successfully reproduce the overall magnetostriction behavior, revealing the competition between the exchange striction and CF striction.
A signature of the CF level crossing is also observed through adiabatic magnetocaloric-effect measurements, consistent with our magnetostriction data.
\end{abstract}

\date{\today}
\maketitle

\section{\label{Sec1}Introduction}
Magnetostriction refers to the distortion of a lattice when a material is subjected to an external magnetic field.
Certain magnetostrictive materials exhibit relative length change $\Delta L/L$ as large as several percent \cite{Hathaway_1993, Yu_2024}.
These materials have been extensively studied for their practical applications in developing multifunctional devices, such as actuators, oscillators, and sensors \cite{Zhang_2012, Valerio_2019}.
Beyond their applicative aspects, magnetostriction serves as a key physical quantity for probing phase transitions as well as underlying spin correlations.
For instance, magnetostriction measurements on the quantum spin ice \PZO{} revealed a sharp first-order transition at extremely low temperatures, shedding light on a liquid-gas metamagnetic transition and offering insights into spin-orbital liquids \cite{NTang_2023, APatri_2020}.
Magnetostriction measurements on the Shastry--Sutherland magnet SrCu$_{2}$(BO$_{3}$)$_{2}$ uncovered a series of spin superstructure phases under ultrahigh magnetic fields \cite{MJaime_2012, Nomura_2023}, which are characterized by devil's-staircase-like magnetization plateaus \cite{YMatsuda_2013}.

\begin{figure}[t]
\centering
\includegraphics[width=\linewidth]{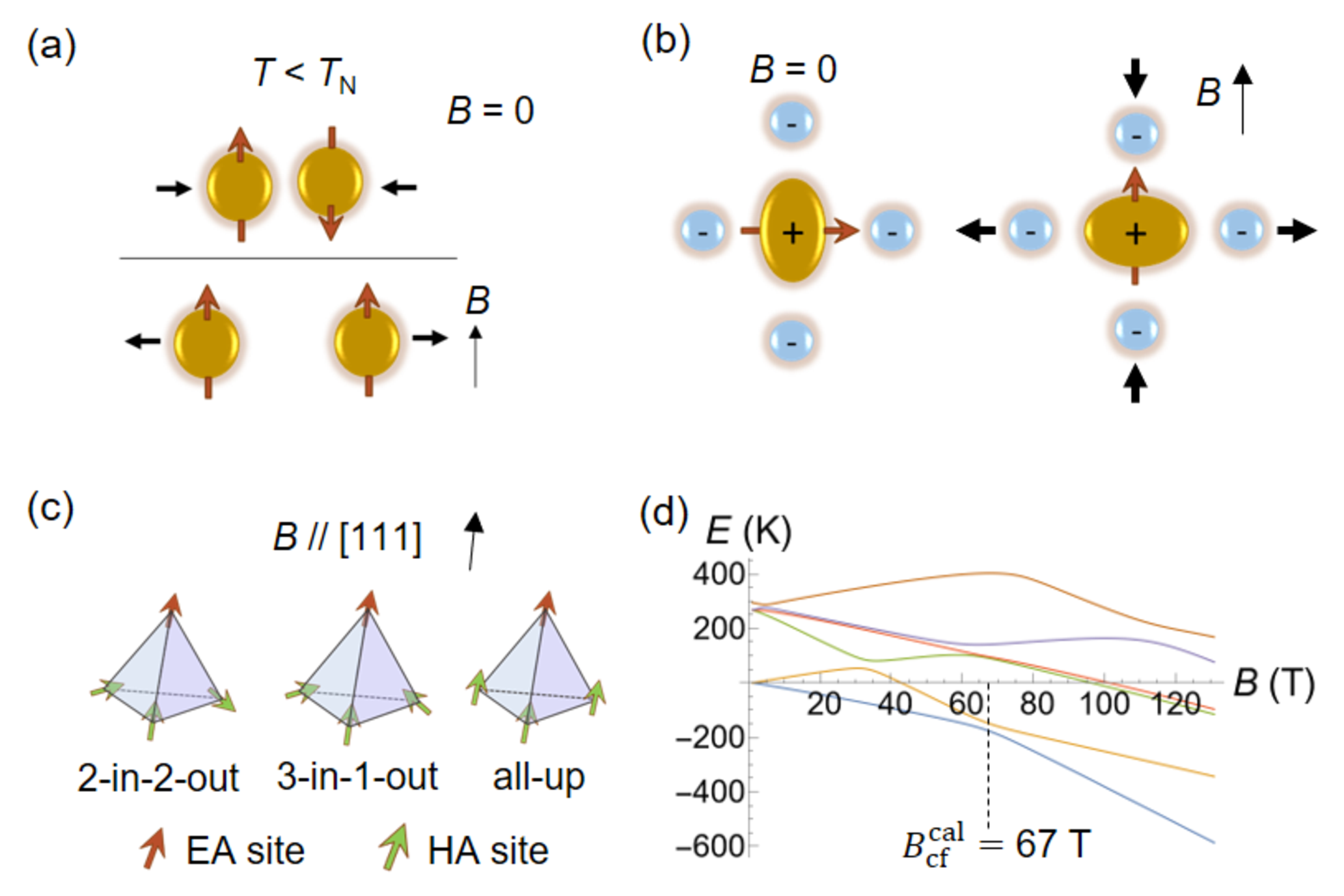}
\caption{(a), (b) Schematics of (a) exchange striction and (b) crystal-field (CF) striction. The orange arrows indicate the magnetic moments, and the yellow (blue) spheres represent the valence electron density of the magnetic ion (ligands). The elliptical deformation of the yellow spheres schematically illustrates the crystal-field anisotropy, which aligns with the magnetic moment direction. The thick black arrows indicate the direction of ion displacement
when the magnetic moment orientation is altered by an external magnetic field $B$. (c) Three types of spin configurations when a magnetic field is applied along [111] on \HTO{}. The spins aligned along the easy axis are colored orange (EA site), while those aligned along the hard axis are colored green (HA site). (d) Energy level diagram for the single-ion Hamiltonian of Eq.~\eqref{eqn_H_singleion}, where one of the three HA spins is subjected to a magnetic field in the [111] direction. The two lowest CF levels hybridize at $B_{\rm cf}^{\rm cal} = 67$~T.}
\label{fig1}
\end{figure}

There are several well-known microscopic origins of magnetostriction \cite{MDoerr_2005}.
One such mechanism involves the dependence of an ionic radius on its electronic configuration.
A field-induced change in the ionic state can result in significant volume magnetostriction, as exemplified by a spin-state transition in LaCoO$_{3}$ \cite{AIkeda_2020} and valence transitions in $f$ electron-based valence-fluctuating materials \cite{KYoshimura_1988, Mus_2004, AMiyake_2022}.
In contrast to these on-site mechanisms, exchange striction refers to the change in distance between two magnetic ions to minimize exchange coupling energy [Fig.~\ref{fig1}(a)]. Here, $\Delta L/L$ reflects changes in local spin correlations $\langle {\bm S}_{i} \cdot {\bm S}_{j} \rangle$ between neighboring sites $i$ and $j$, and consequently correlates with the magnetization $M$ \cite{Zapf_2008, MJaime_2012, AIkeda_2019, AMiyata_2021}.
The crystal field (CF) also influences the arrangement of surrounding anions and the wavefunction of the magnetic site to minimize Coulomb energy.
Applying a magnetic field alters the wavefunction through CF level hybridization, potentially causing anisotropic lattice deformation, known as CF striction [Fig.~\ref{fig1}(b)].
Although magnetostriction is universally observed in all magnetic materials, its theoretical description is far from straightforward and is often simplified by phenomenological approaches \cite{Patri_2019} or effective magnetoelastic models \cite{Kimura_2014, AMiyata_2020}.

The target compound in this study is \HTO{}, one of the most extensively studied $4f$ Ising pyrochlore magnet, known as a classical spin ice \cite{Harris_1997, ROsenkranz_2000, KMatsuhira_2000, STBramwell_2001, Petrenko_2003, Fennell_2009, YNakanishi_2011, CCastelnovo_2012, SErfanifam_2014, MRuminyDFT_2016, MRuminy_2016_CF, LOpherden_2019, Wang_2021}.
The strong Ising anisotropy along the $\langle 111 \rangle$ axes arises from a combination of dominant nearest-neighbor ferromagnetic interactions and the CF effects. This leads to a highly degenerate ground state characterized by short-range spin correlations, where two spins point inward and two point outward within each tetrahedron (``2-in--2-out") \cite{Petrenko_2003} [Fig.~\ref{fig1}(c)].
When a magnetic field is applied along the [111] direction, a metamagnetic transition occurs at approximately 2~T, flipping one spin and resulting in a ``3-in--1-out'' (or ``1-in--3-out'') configuration in the tetrahedra, representing a fully polarized state within the Ising limit [Fig.~\ref{fig1}(c)].
In this field orientation, each tetrahedron has one magnetic site with the easy axis parallel to $B$ [the easy-axis (EA) site] and three sites with axes not parallel to $B$ [the hard-axis (HA) sites].

Given the strong spin-orbit coupling that links spin and orbital degrees of freedom, the CF-phonon interaction plays a key role in magnetoelastic properties, forming the foundation of magnetostrictive phenomena.
However, as the easy-axis directions vary at the four magnetic sites in  \HTO{}, local lattice deformations induced by the CF tend to cancel out, making its effect on the crystal structure less evident.
To clarify this, the observation of the CF striction in high magnetic fields would be useful.
In \HTO{}, the energy gap between the lowest and the first-excited CF levels is approximately 260~K in zero field \cite{ROsenkranz_2000, MRuminy_2016_CF}.
The application of an external magnetic field along [111] can hybridize these CF levels at the HA sites, as shown in the CF level diagram in Fig.~\ref{fig1}(d), eventually realizing a forced ferromagnetic (all-up) state [Fig.~\ref{fig1}(c)].
The CF level crossing is theoretically expected at $B_{\rm cf}^{\rm cal} = 67$~T based on the exact diagonalization of the CF Hamiltonian.
Indeed, Erfanifam {\it et al.} observed a signature of the CF level hybridization for $B \parallel [111]$ through magnetization and ultrasound measurements up to 60~T \cite{SErfanifam_2014}, and Opherden {\it et al.} observed a sequence of metamagnetic transitions accompanied by the CF level crossing for $B \parallel [5513]$ up to 120~T \cite{LOpherden_2019}.

Here, we investigate the CF striction in \HTO{} using magnetostriction measurements under ultrahigh magnetic fields up to 120~T, complemented by magnetization and magnetocaloric effect (MCE) measurements.
We observe significant anisotropic CF striction, characterized by a lattice contraction amounting to $\Delta L/L \sim -5 \times 10^{-4}$ along the field direction.
We were also able to resolve the timescale of spin dynamics, finding it to be comparable to the microsecond range, relevant to monopole formation and annihilation. Additionally, CF-phonon dynamics appear to occur at rates that are comparable to, or potentially even faster than, the microsecond range.
Our mean-field calculations, based on a point-charge model and incorporating CF, two-ion exchange interactions, phonons, and CF-phonon coupling, successfully reproduce the magnetostriction data.
Similar high-field magnetostriction is observed in another spin ice compound, \PZO{}, suggesting that the CF striction is a common feature in rare-earth-based spin-ice systems.

\vspace{-0.2cm}
\section{\label{Sec2}Methods}

\vspace{-0.4cm}
\subsection{\label{Sec2A}Experiments}

Single crystals of \HTO{} and \PZO{} were grown from polycrystalline feed rods using a high-temperature xenon-type optical floating zone furnace at ISSP, University of Tokyo. Crystal orientations were checked using a backscattering Laue x-ray diffractometer.
Several pieces of as-grown crystals were cut and polished into rectangular parallelepipeds with dimensions of approximately 1 $\times$ 1 $\times$ 3 mm$^3$.

\begin{table*}[t]
\centering
\renewcommand{\arraystretch}{1.2}
\caption{Experimental conditions for magnetization $M$, magnetostriction $\Delta L/L$, and magnetocaloric effect (MCE) measurements.}
\begin{tabular}{ccccc} \hline\hline
~~ & ~Pulsed magnet~ & ~Field duration~ & ~Maximum field~ & ~Temperature conditions~ \\ \hline
~~$M$~~ & \parbox{4.0cm}{\centering \vspace{+0.1cm}HSTC \\ \centering \vspace{+0.05cm}Nondestructive short pulse \vspace{+0.05cm}} & \parbox{2.5cm}{\centering \vspace{+0.05cm}\hspace{+0.25cm}8~$\mu$s \\ \centering \vspace{+0.05cm}\hspace{+0.25cm}4~ms \vspace{+0.05cm}} & \parbox{2.5cm}{\centering \vspace{+0.05cm}130~T \\ \centering \vspace{+0.05cm}\hspace{+0.24cm}62~T \vspace{+0.05cm}} & \parbox{4.0cm}{\centering \vspace{-0.04cm}Quasiadiabatic\\ \vspace{+0.05cm}Nonadiabatic} \\ \hline
~~$\Delta L/L$~~ & \parbox{4.0cm}{\centering \vspace{+0.1cm}HSTC \\ \centering \vspace{+0.05cm}Nondestructive long pulse \vspace{+0.05cm}} & \parbox{2.5cm}{\centering \vspace{+0.05cm}\hspace{+0.24cm}8~$\mu$s \\ \centering \vspace{+0.05cm}\hspace{+0.08cm}36~ms \vspace{+0.05cm}} & \parbox{2.5cm}{\centering \vspace{+0.05cm}120~T \\ \centering \vspace{+0.05cm}\hspace{+0.24cm}55~T \vspace{+0.05cm}} & \parbox{4.0cm}{\centering \vspace{-0.04cm}Quasiadiabatic\\ \vspace{+0.05cm}Nonadiabatic} \\ \hline
~~MCE~~ & ~\hspace{-0.08cm}Nondestructive long pulse~ & ~36~ms~ & ~\hspace{+0.16cm}55~T~ & ~~Quasiadiabatic \& nonadiabatic~~ \\ \hline\hline
\end{tabular}
\label{tab1}
\end{table*}

We performed magnetization and magnetostriction measurements in the field orientation $B \parallel [111]$.
Magnetization up to 7~T was measured using a superconducting quantum interference device (MPMS, Quantum Design).
Magnetization up to 62~T and 130~T was measured by the induction method in a nondestructive short-pulsed magnet (4~ms duration) and in a horizontal single-turn-coil (HSTC) system (8~$\mu$s duration) \cite{MGen_2020}, respectively.
Longitudinal magnetostriction ($\Delta L/L \parallel [111]$) up to 55~T and 118~T was measured by the fiber-Bragg-grating (FBG) method in a nondestructive long-pulsed magnet (36~ms duration) and in the HSTC system, respectively.
Transverse magnetostriction ($\Delta L/L \perp$ [111]) up to 55~T was also measured in the nondestructive long-pulsed magnet, where the optical fiber was bent by 90 degrees, as in Refs.~\cite{AMiyake_2022, MGen_2022}.
A relative sample-length change $\Delta L/L$ was detected by the optical filter method \cite{AIkeda_2017}.
The optical fiber was glued on the flat surface of the crystal using Stycast1266.

In the nondestructive pulsed magnet, the magnetization and magnetostriction measurements were performed in either liquid-$^4$He or gas $^4$He environment, i.e., nonadiabatic conditions.
In the HSTC system, the sample was cooled to approximately 5~K using a liquid-$^4$He flow-type cryostat, where the measurement condition should be quasiadiabatic because of the short pulsed-field duration.
To monitor the temperature change of the sample during the field sweep, we measured the magnetocaloric effect (MCE) up to 55~T under both nonadiabatic and quasiadiabatic conditions using the nondestructive long-pulsed magnet \cite{2013_Kih}.
A sensitive Au$_{16}$Ge$_{84}$ film thermometer was sputtered on the surface of the crystal, with temperature calibration performed using a commercial RuO$_{2}$ thermometer.

Table~\ref{tab1} summarizes the measurement conditions for all experiments conducted under pulsed magnetic fields.
The typical magnetic-field profiles of the three types of magnets used in the experiments are shown in Fig.~\ref{fig7} in Appendix~\ref{AppendixA}.

\vspace{-0.4cm}
\subsection{\label{Sec2B}Theoretical calculations}

To capture the CF level scheme in \HTO{}, we first consider a single-ion Hamiltonian that includes both the CF and Zeeman terms. The CF Hamiltonian can be expressed as follows:
\begin{equation}
\begin{aligned}
\label{eq_H_cf} 
\hat{H}_{\rm cf} =& \sum_{lm} B_{lm} O_{lm} \\
\end{aligned}
\end{equation}
where $B_{lm}$ denotes the CF parameters, and $O_{lm}$ represents Stevens Operator equivalents.
The single-ion Hamiltonian is then written as
\begin{equation}
\label{eqn_H_singleion}
\hat{H}_{\rm single-ion} = \hat{H}_{\rm cf}-g_{J}\mu_{\rm B}\hat{\bm J}_{i} \cdot {\bm B},
\end{equation}
where $g_{J}$ is the Land\'{e}'s $g$ factor, $\mu_{\rm B}$ is the Bohr magneton, $\hat{\bm J}_{i}$ is the magnetic moment at site $i$, and ${\bm B}$ is the external magnetic field.
We obtain the field evolution of the CF level energy at the HA site [Fig.~\ref{fig1}(d)] by exact diagonalization of eq.~\eqref{eqn_H_singleion} \cite{notes}.

For detailed analyses of magnetization and magnetostriction, we additionally take into account magnetic interactions, phonons, and CF-phonon coupling.
The magnetic properties of \HTO{} can be accurately described by the classical dipolar spin ice model, incorporating dominant long-range dipolar interaction $J_{\rm dipo}$ and short-range nearest-neighbor exchange interaction $J_{\rm ex}$.
As the dipolar interaction $J_{\rm dipo}$ can be effectively renormalized into the nearest-neighbor exchange interaction \cite{STBramwell_2001}, we consider a two-spin exchange coupling term described by
\begin{align}
\label{eq_H_ex}
\hat{H}_{\rm ex} &=-\frac{1}{2} {\sum_{ij} J(\bm{R}_{ij})\hat{\bm J}_{i}\hat{\bm J}_{j}},
\end{align}
where the sum is taken over all nearest-neighbor pairs, $J(\bm{R}_{ij}) \equiv J_{\rm dipo} + J_{\rm ex}$ is the distance-dependent effective exchange coupling, and $\bm{R}_{ij}$ is the vector connecting sites $i$ and $j$.
Assuming that $J$ is linearly modulated by a small strain $\epsilon$, we expand $J({\bm{R}}_{ij})$ as,
\begin{align}
\label{exchange_const}
J (\bm{R}_{ij}) \approx J (\bm{R}_{ij}^{0}) + \sum _{\alpha, \gamma=1,2,3, \beta = 1, ..., 6}\frac{\partial J}{\partial R^\alpha}\frac{\partial \epsilon_{\alpha \gamma}R_{ij}^\gamma}{\partial \epsilon_\beta} \epsilon_\beta,
\end{align}
where $\bm{R}_{ij}^{0}$ is the distance between the original positions of sites $i$ and $j$, and the indices $\alpha, \gamma = x, y, z$ refer to an Euclidean coordinate system parallel to the crystal axes, i.e., $x \parallel a$, $y \parallel b$ and $z \parallel c$. The index $\beta$ follows Voigt notation, ranging from 1 - 6. In other words, $\epsilon_\beta$ represents the strain tensor $\epsilon_{\alpha \gamma}$ in Voigt notation. 
For the lattice contribution, we consider a phonon term, $\hat{H}_{\rm ph}$, and a CF-phonon term, $\hat{H}_{\rm cfph}$, which represents the coupling between the CF and phonons.
The detailed formulas for these two terms are provided in Appendix~\ref{AppendixB}.
Combining all these terms, we construct the total Hamiltonian as
\begin{equation}
\label{eqn_H_total}
\hat{H}_{\rm total} =  \sum _{i}{\hat{H}_{\rm single-ion}} + \hat{H}_{\rm ex} + \hat{H}_{\rm ph} + \hat{H}_{\rm cfph}.
\end{equation}

We performed numerical simulations based on Eq.~\eqref{eqn_H_total} using the \textsf{McPhase} program \cite{MRotter_2002, MRotter_2004, MRotter_2012}, which specializes in calculating the physical properties of $4f$-based magnets within the framework of mean-field theory \cite{MRotter_2007, JJensen_2007, TStoter_2020}. 
\textsf{McPhase} employs a point-charge model to determine the CF wavefunctions and CF energy gap \cite{notes}.
To match the experimental energy gap of approximately 260 K \cite{ROsenkranz_2000, MRuminy_2016_CF}, we apply a scaling factor of 0.7 to the nominal charges Ho ($3+$), Ti ($4+$), O ($2-$). To set the initial state in zero field, we utilized the crystallographic parameters of \HTO{} reported in Ref.~\cite{Baroudi_2015}.
We set the dipolar interaction to $J_{\rm dipo} = 2.4$~K \cite{STBramwell_2001} and the nearest-neighbor exchange interaction to $J_{\rm ex} = 0.1$~K, yielding a total effective exchange coupling of $J = 2.5$~K.
This set of parameters successfully reproduces both the metamagnetic transition in the low-field regime and the CF level crossing at $B_{\rm cf} \sim 65$ T, as experimentally observed.
We note that even a small change of 0.1~K in $J_{\rm ex}$ results in a 6~T shift in $B_{\rm cf}^{\rm cal}$.
In Eq.~\eqref{exchange_const}, we assume $dJ/dR \simeq dJ_{\rm ex}/dR$, which is determined by the Bethe-Slater curve, as shown in Fig.~\ref{fig8} in Appendix~\ref{AppendixB}.

To calculate the magnetization, we solve the mean-field equations self-consistently to obtain the magnetization at each sublattice $i$, which is given by
\begin{equation}
\label{eqn_M}
\bm {M}_i = \sum_{n} \frac{g_J \mu_{\rm B} \langle n |\hat{\bm J_i}| n \rangle}{Z} \exp(-E_n/k_{\rm B}T),
\end{equation}
where $Z = \sum_{n}\exp(-E_n/k_{\rm B}T)$ is the partition function for sublattice $i$, and $\vert n\rangle$ is the eigenstate corresponding to the eigenvalue $E_{n}$ of $\hat{H}_{\rm total}$ treated in mean field approximation. The total magnetization $M$ is obtained by averaging the magnetization of all sublattices, in the unit of $\mu_{\rm B}$/Ho.

For the magnetostriction, we calculate strains self-consistently using Eq.~\eqref{eqn_H_total} and the following equation:
\begin{equation}
\begin{aligned}
\label{eqn_strain_selfconsistent}
\sum_{\beta=1,...,6}c^{\alpha \beta}\epsilon_{\rm \beta} 
&=\sum_{i,\delta=1,2,3}G^{\alpha \delta}_{\rm mix}(i)\langle u^{\delta}_i \rangle \\
&+ \sum_{i,lm}G_{\rm cfph}^{\alpha lm}(i)\langle O_{lm}(\hat{\bm J_i})\rangle \\
&+ \frac{1}{2}\sum_{i,i',\delta,\delta',\alpha',\gamma=1, 2, 3} \frac{\partial J_{\delta \delta'} (\bm R_{ii'})}{\partial R^{\alpha'}}\frac{\partial \epsilon_{\alpha ' \gamma} \partial R^\gamma_{ii'}} {\partial \epsilon_{\alpha}} \langle \hat{J}_{i\delta}\hat{J}_{i'\delta'} \rangle,
\end{aligned}
\end{equation}
where $i$, $i'$, $\delta$, $\delta'$, $\alpha'$, and $\gamma$ run over the spatial coordinates $x$, $y$, and $z$, represented as the integers 1, 2, and 3, respectively.
$c^{\alpha \beta}$ denotes the elastic constant, where $\alpha$ and $\beta$ are indices in Voigt notaion, ranging from 1 - 6.
$G_{\rm mix}^{\alpha \delta}$ represents the phonon-strain coupling, and $G_{\rm cfph}^{\alpha lm}$ represents the CF-phonon coupling.
For details of these parameters, see Appendix~\ref{AppendixB}.
In Eq.~(\ref{eqn_strain_selfconsistent}), the first and second terms on the right side contribute to the CF striction, and the third term is the exchange striction.
The total lattice strain, i.e., magnetostriction, can be obtained by
\begin{equation}
\label{eqn_dL}
\frac {\Delta L}{L} = \sum_{i,j=1,2,3}\epsilon_{ij}{\bm{{l}}_{i}}{\bm{l}_{j}},
\end{equation}
where $\bm{l}$ denotes the unit vector in the direction of measurement.
Strain $\epsilon_{ij}$ is a $3 \times 3$ symmetric tensor, hence has six independent matrix elements: $\epsilon_{xx}$, $\epsilon_{yy}$, $\epsilon_{zz}$, $\epsilon_{xy}$,  $\epsilon_{xz}$,  $\epsilon_{yz}$.
We obtain the longitudinal magnetostriction ($\Delta L/L \parallel$ [111]) by calculating $\Delta L/L=(\epsilon_{xx}+\epsilon_{yy}+\epsilon_{zz}+ 2(\epsilon_{xy}+\epsilon_{xz}+\epsilon_{yz}))/3$. In this study, we use Voigt notation, where $\epsilon_{ij}$ is relabeled as $\epsilon_{1}=\epsilon_{xx}$, $\epsilon_{2}=\epsilon_{yy}$, $\epsilon_{3}=\epsilon_{zz}$, $\epsilon_4=2\epsilon_{yz}=2\epsilon_{zy}$,  $\epsilon_5=2\epsilon_{xz}=2\epsilon_{zx}$, and $\epsilon_6=2\epsilon_{xy}=2\epsilon_{yx}$.

\vspace{-0.2cm}
\section{\label{Sec3}Results and Discussion}

\vspace{-0.4cm}
\subsection{\label{Sec3A}High-field magnetization and magnetostriction}

\begin{figure}[t]
\centering
\includegraphics[width=0.9\linewidth]{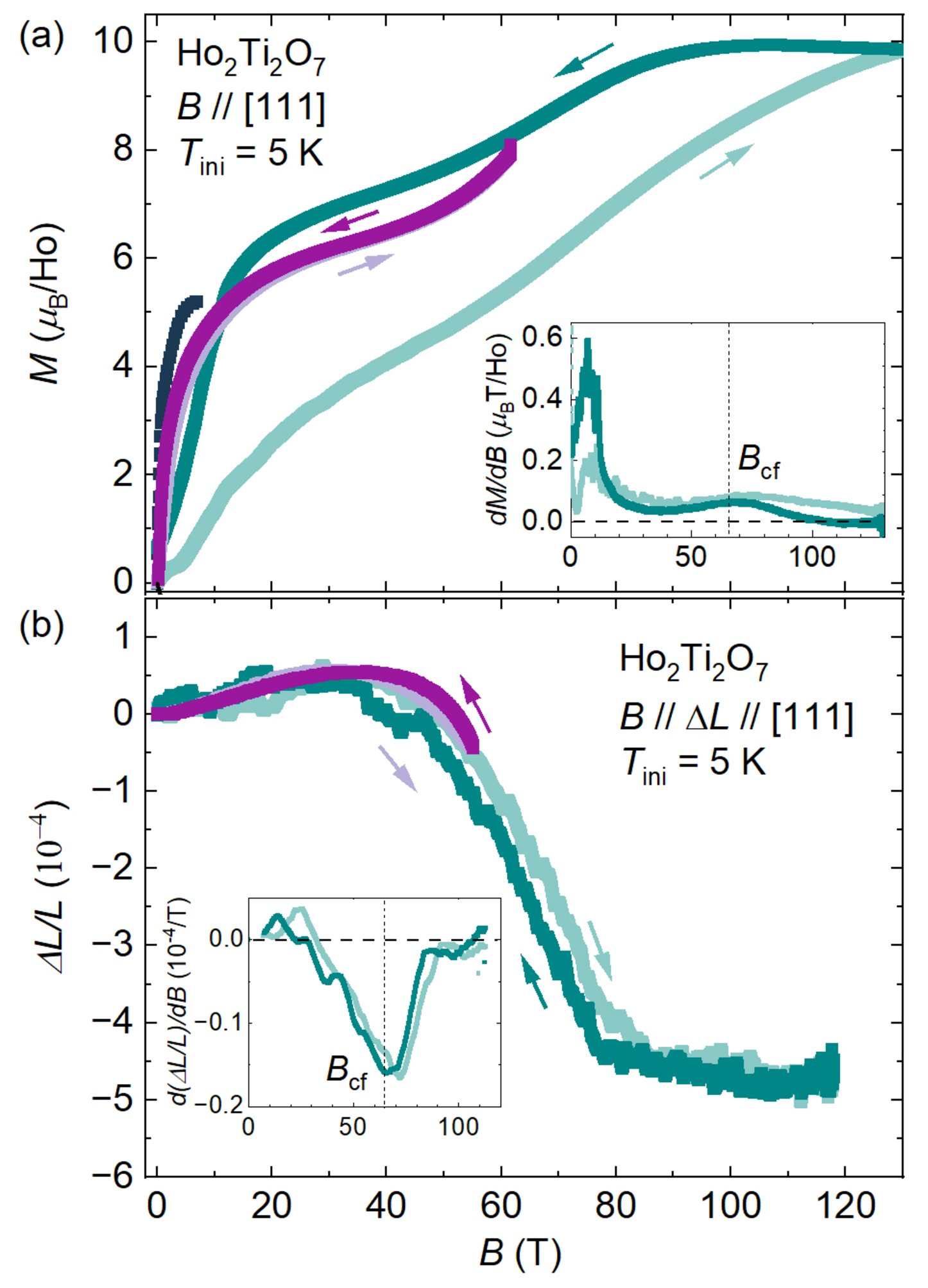}
\caption{Magnetic-field dependence of (a) magnetization $M$ and (b) longitudinal magnetostriction $\Delta L/L \parallel [111]$ in \HTO{} for $B \parallel [111]$ in ultrahigh magnetic fields. Black symbols in (a) show the $M$--$B$ data measured at $T = 4.2$~K using the MPMS, purple symbols show the data measured at $T_{\rm ini} = 4.2$~K using the nondestructive pulsed magnet, and green symbols show the data measured at $T_{\rm ini} = 5$~K using the HSTC system. The insets of (a) and (b) show the field derivatives of magnetization $dM/dB$ and magnetostriction $d(\Delta L/L)/dB$, respectively, for the HSTC data. The critical field of the CF level crossing $B_{\rm cf}$ is defined using the field-down sweep data.}
\label{fig2}
\end{figure}

Figure~\ref{fig2}(a) shows the magnetization curve of \HTO{} for $B \parallel [111]$ measured up to 130~T at an initial temperature of $T_{\rm ini} = 5$~K using the HSTC system. The $M$--$B$ curve measured up to 62~T at $T_{\rm ini} = 4.2$~K using the nondestructive pulsed magnet is also displayed, which reproduces Ref.~\cite{SErfanifam_2014}. Both $M$--$B$ curves exhibit a plateau-like feature between 20 and 60~T, followed by a metamagnetic increase, where the field derivative of magnetization $dM/dB$ exhibits a broad peak at $B_{\rm cf} \sim 65$~T for the field-down sweep [see the inset of Fig.~\ref{fig2}(a)].
This metamagnetic behavior suggests the CF level crossing at the HA sites, consistent with the CF scheme shown in Fig.~\ref{fig1}(d). Above $B_{\rm cf}$, $M$ gradually approaches the expected full moment, 10~$\mu_{\rm B}$/Ho, suggesting the realization of the ``all-up'' state.
We note that the sensitivity of magnetization detection decreases as $dB/dt$ approaches zero, which may introduce errors in the absolute value of $M$ near the maximum field.

Figure~\ref{fig2}(b) shows the longitudinal magnetostriction curve for $B \parallel [111]$ measured up to 120~T at $T_{\rm ini} = 5$~K using the HSTC system, along with the curve measured up to 55~T at $T_{\rm ini} = 4.2$~K using the nondestructive pulsed magnet. Initially, $\Delta L/L$ increases and reaches a maximum around 30~T. Subsequently, $\Delta L/L$ reverses and dramatically decreases from 30 to 80~T, where the field derivative of the magnetostriction $d(\Delta L/L)/dB$ exhibits a peak at $B_{\rm cf} \sim 65$~T for the field-down sweep [see the inset of Fig.~\ref{fig2}(b)].
Finally, $\Delta L/L$ is nearly saturated at $\sim$80~T.
The observed change in sample-length across $B_{\rm cf}$, amounting to $\Delta L/L \sim 5 \times 10^{-4}$, highlights the strong CF effect on the crystal structure of \HTO{}.

\vspace{-0.4cm}
\subsection{\label{Sec3B}Magnetocaloric effect (MCE)}

\begin{figure}[t]
\centering
\includegraphics[width=0.8\linewidth]{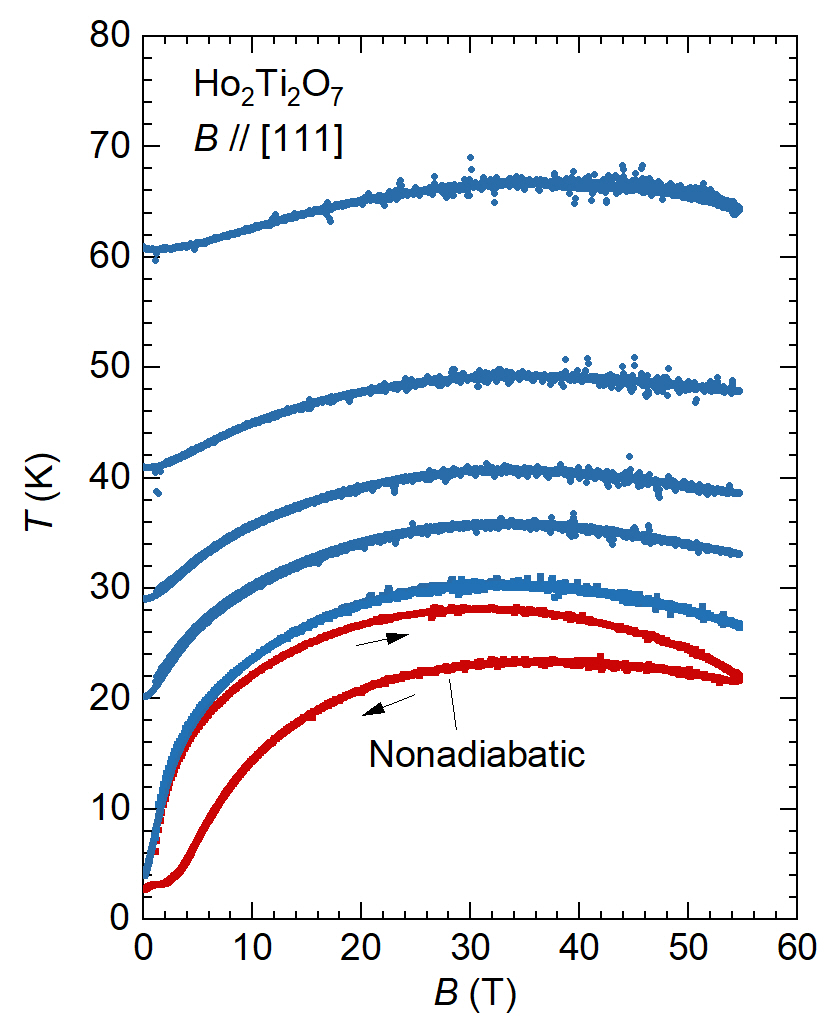}
\caption{Magnetocaloric effect in \HTO{} for $B \parallel [111]$ measured at various $T_{\rm ini}$'s under quasiadiabatic conditions (blue), and at $T_{\rm ini} = 4.2$~K under nonadiabatic conditions (red).}
\label{fig3}
\end{figure}

Comparing the three $M$--$B$ curves shown in Fig.~\ref{fig2}(a), it is evident that the initial rise in magnetization at low magnetic fields becomes more gradual as the field sweep becomes faster.
This trend could be attributed to two factors: sample heating and slow spin relaxation.
We will discuss the spin dynamics in Sec.~\ref{Sec3C}.
To verify the sample heating, we performed MCE measurements for $B \parallel [111]$ up to 55~T in the nondestructive pulsed magnet.
The field dependence of the sample temperature $T(B)$, measured at various initial temperatures ($T_{\rm ini}$'s) under quasiadiabatic conditions, are shown by the blue curves in Fig.~\ref{fig3}.
The $T(B)$ curves for field-up and field-down sweeps overlap for all the measured $T_{\rm ini}$'s.
The absence of hysteresis ensures that nearly adiabatic conditions were achieved.
These $T(B)$ curves can hence be regarded as isentropic, meaning that the total entropy, comprising both lattice entropy and magnetic entropy, remains constant.

For $T_{\rm ini} = 5$~K, $T(B)$ approaches 30~K at 30~T, indicating a significant reduction in magnetic entropy \cite{Kittaka_2018}.
This occurs because the highly degenerate 2-in--2-out states are destroyed by a magnetic field as weak as 2~T, stabilizing the nondegenerate 3(1)-in--1(3)-out state. At higher initial temperatures, the increase in $T(B)$ is less pronounced due to greater thermal fluctuations and larger lattice heat capacity.
After $T(B)$ reaches its maximum around 35~T, the sample gradually cools as the magnetic field increases up to 55~T.
This cooling behavior indicates an increase in magnetic entropy, suggesting the onset of CF level hybridization.

We also measured the sample temperature changes at $T_{\rm ini} = 4.2$~K in a liquid-$^{4}$He environment, i.e., under nonadiabatic conditions, as shown by the red curve in Fig.~\ref{fig3}.
Surprisingly, $T(B)$ reaches 28~K during the field-up sweep, which is only slightly lower than in the quasiadiabatic case.
These MCE results indicate that sample heating effects are unavoidable when measuring \HTO{} in pulsed high magnetic fields.
We infer that, in the magnetization and magnetostriction data presented in Fig.~\ref{fig2}, the sample temperature is likely between 20 and 30~K at fields above 10~T.

\vspace{-0.4cm}
\subsection{\label{Sec3C}Spin dynamics}

Our high-field data also provide valuable insights into the dynamic properties of the magnetic state.
In the experimental data obtained using the millisecond-duration pulsed magnetic fields [Figs.~\ref{fig7}(a) and \ref{fig7}(b)], both the $M$--$B$ curve up to 62~T and the $\Delta L/L$--$B$ curve up to 55~T exhibit small hysteresis (Fig.~\ref{fig2}).
The hysteresis is likely caused by a lower sample temperature during the field-down sweep compared to the field-up sweep, as suggested by our nonadiabatic MCE data (Fig.~\ref{fig3}). We note that the hysteresis loop in the $\Delta L/L$--$B$ curve cannot be attributed to slow spin dynamics because the sign of the hysteresis is opposite to what would be expected from a delayed response. Therefore, the small hysteresis observed in millisecond-duration pulsed magnetic-fields suggests the timescale of spin relaxation is faster than the millisecond range, which will be discussed in more detail in the next paragraph.

\begin{figure}[t]
\centering
\includegraphics[width=0.7\linewidth]{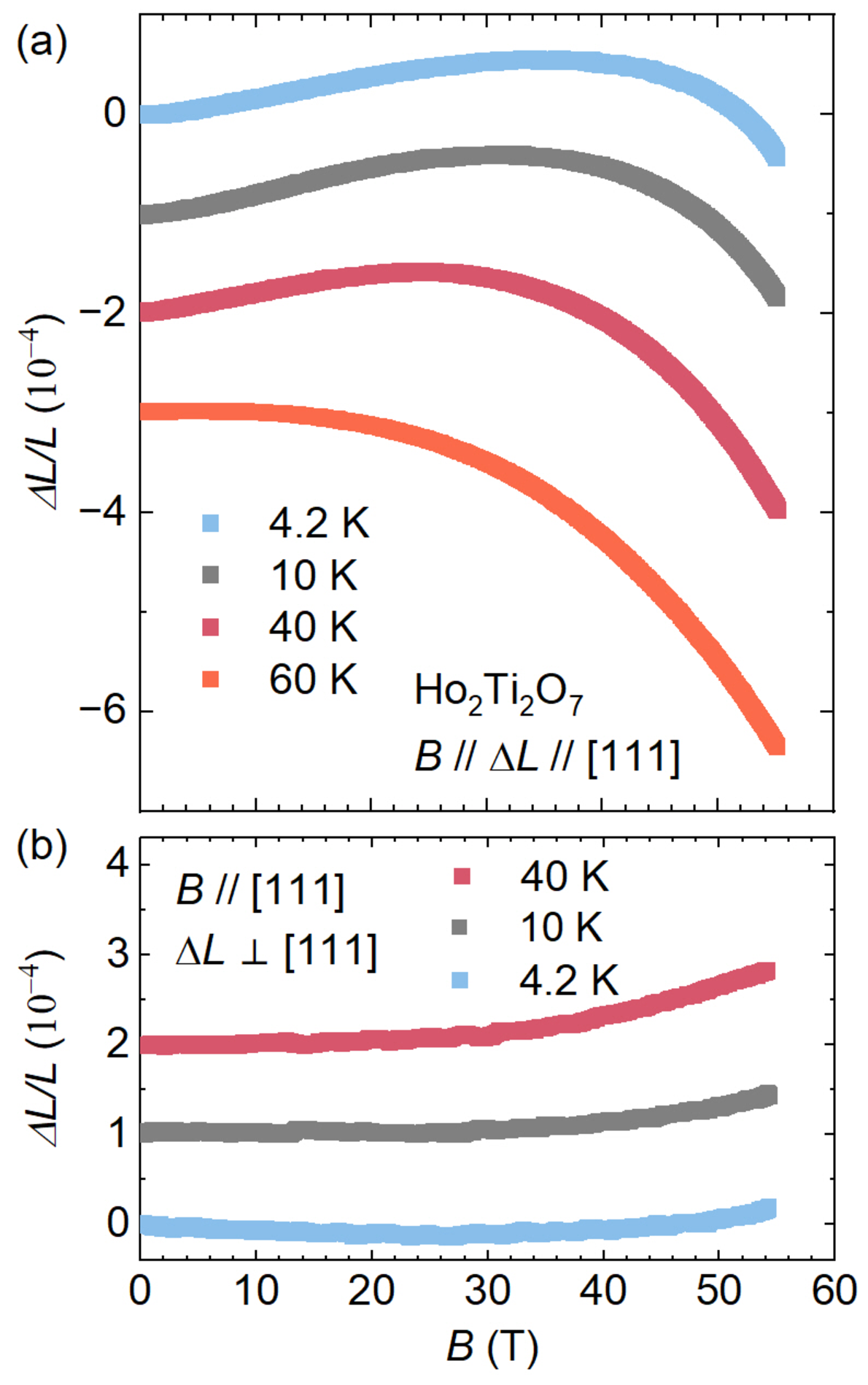}
\caption{Temperature dependence of
(a) longitudinal and (b) transverse magnetostriction in Ho$_{2}$Ti$_{2}$O$_{7}$ for $B \parallel [111]$. The data except for $T_{\rm ini} = 4.2$~K are offset for clarity.}
\label{fig4}
\end{figure}

In contrast, the $M$--$B$ curve up to 130~T, obtained in the microsecond-duration pulsed magnetic fields [Fig.~\ref{fig7}(c)] under quasi-adibatic condition, exhibits significant hysteresis [Fig.~\ref{fig2}(a)].
Given that no hysteresis is observed in the quasiadiabatic MCE (Fig.~\ref{fig3}), the observed hysteresis can be attributed to slow spin relaxation, which mainly originates from nonequilibrium processes of monopole formation or annihilation \cite{TStoter_2020}.
Together with the magnetization data from millisecond-pulsed fields, the spin dynamics appears to be faster than the millisecond but slower than or comparable to the microsecond regime, consistent with previous studies based on neutron spin echo and ac susceptibility experiments \cite{GEhlers_2003}. However, inelastic neutron scattering experiments have reported a different timescale of 10 ns, suggesting the presence of faster processes \cite{JPClancy_2009}. The reason for such discrepancy has been addressed in Ref. \cite{Quilliam_2011}.

Notably, the hysteresis observed in the $\Delta L/L$--$B$ curve up to 120~T is much less pronounced compared to that in the $M$--$B$ curve (Fig.~\ref{fig2}), a difference that cannot be explained by temperature effects, as both experiments were performed under quasiadiabatic condition. The overall sample-length change of \HTO{} in high magnetic fields primarily originates from CF-striction, with a smaller contribution from exchange striction (see Sec.~\ref{Sec3E}), making CF-striction the main source of hysteresis. This suggests that the dynamics of CF-striction is likely within or faster than the microsecond range. 
It is important to note that potential contributions from slower spin-lattice dynamics cannot be excluded, given the absence, to the best of our knowledge, of previous research on such interactions in \HTO{}, as these dynamics should occur beyond the timescale of our pulsed-field experiments. Further investigation into the dynamics of spin-lattice interactions in \HTO{} would be valuable for future work.

\vspace{-0.4cm}
\subsection{\label{Sec3D}Temperature dependence and anisotropic magnetostriction}

We also investigate the anisotropy and temperature dependence of the magnetostriction.
Figures~\ref{fig4}(a) and \ref{fig4}(b) show the longitudinal and transverse magnetostriction of \HTO{}, respectively, for $B \parallel [111]$ measured at various $T_{\rm ini}$'s in the nondestructive pulsed magnet.
As $T_{\rm ini}$ increases, the position of the hump in the longitudinal magnetostriction curve shifts to a lower field, and eventually, the hump disappears at $T_{\rm ini} = 60$~K.

In contrast to the longitudinal magnetostriction, the transverse magnetostriction at $T_{\rm ini} = 4.2$~K exhibits a broad dip around 30~T.
As the temperature increases to $T_{\rm ini} = 40$~K, this dip structure disappears, and $\Delta L/L$ begins to increase steadily with the magnetic field.
The anisotropic behavior between the longitudinal and transverse magnetostricition at high magnetic fields is the characteristic of the CF striction.

\vspace{-0.4cm}
\subsection{\label{Sec3E}Numerical calculations}

\begin{figure}[t]
\centering
\includegraphics[keepaspectratio, scale=0.32]{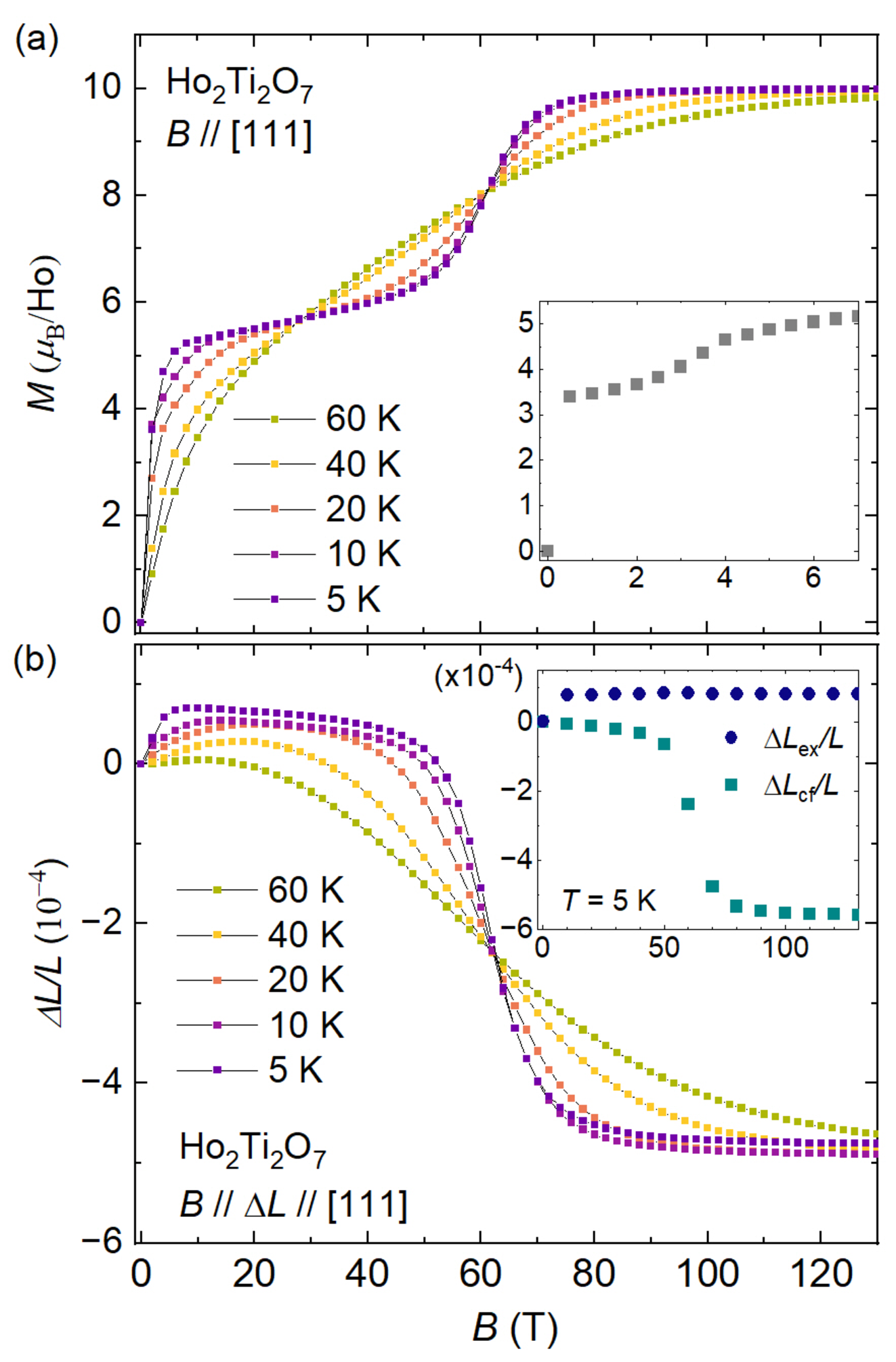}
\caption{Calculation results of (a) $M$--$B$ curves and (b) $\Delta L/L$--$B$ curves for the Hamiltonian in Eq.~\eqref{eqn_H_total}, obtained by mean-field approach. The inset of (a) shows an enlarged view of the $M$--$B$ curve at 5~K in the low-field regime, clarifying a metamagnetic behavior from the ``2-in--2-out" to the ``3(1)-in--1(3)-out" state. In the inset of (b), the circle and square symbols represent the contributions from exchange striction ($\Delta L_{\rm ex}/L$) and crystal-field (CF) striction ($\Delta L_{\rm cf}/L$), respectively, to the total magnetostriction at 5~K.}
\label{fig5}
\end{figure}

To theoretically understand the behavior of $M$ and $\Delta L/L$ in high magnetic fields, we performed numerical calculations based on the mean-field approach for the Hamiltonian in Eq.~\eqref{eqn_H_total}.
Figures~\ref{fig5}(a) and \ref{fig5}(b) show the calculated $M$--$B$ and $\Delta L/L$--$B$ curves, respectively, for $B \parallel [111]$ at various temperatures.
By adopting the ``2-in--2-out'' spin-ice structure as the initial state in zero field, the metamagnetic transition to the ``3(1)-in--1(3)-out'' state is reproduced at 5~K, as shown in the inset of Fig.~\ref{fig5}(a).
At low temperatures below 20~K, a plateau-like feature appears between 10 and 50~T, followed by another metamagnetic transition at approximately 65~T induced by the CF level crossing.
These trends are qualitatively consistent with the experimentally observed $M$--$B$ curve [Fig.~\ref{fig2}(a)].
The less pronounced plateau-like feature and broader metamagnetic behavior in the experimental curve can be attributed to the MCE and slow spin relaxation, as mentioned in Secs.~\ref{Sec3B} and \ref{Sec3C}.

Having Confirmed that our simulations accurately describe the magnetization of \HTO{} across the entire field range, we now turn to the longitudinal magnetostriction.
According to Eq.~\eqref{eqn_strain_selfconsistent}, the total magnetostriction is composed of the CF striction ($\Delta L_{\rm cf}/L$) and exchange striction ($\Delta L_{\rm ex}/L$).
The field dependence of $\Delta L_{\rm cf}/L$ and $\Delta L_{\rm ex}/L$ at 5~K is plotted in the inset of Fig.~\ref{fig5}(b), where $\Delta L_{\rm cf}/L$ becomes more prominent in the high-field regime.
The combination of a positive change in $\Delta L_{\rm ex}/L$ in the low-field regime and a negative change in $\Delta L_{\rm cf}/L$ in the high-field regime results in a hump structure centering around 30 T in the total magnetostriction curve.
Notably, the rounded shape of the experimentally observed hump in the $\Delta L/L$--$B$ curve at $T_{\rm ini} = 5$~K [Fig.~\ref{fig2}(b)] is presumably due to sample heating, which is consistent with the calculated $\Delta L/L$--$B$ curve at 20~K.
Our calculations predict that the hump in $\Delta L/L$ shifts to a lower magnetic field with increasing temperature and eventually disappears above 60~K.
This behavior also agrees well with the experimental observations, as shown in Fig.~\ref{fig4}(a).

\begin{figure}[t]
\centering
\includegraphics[width=0.8\linewidth]{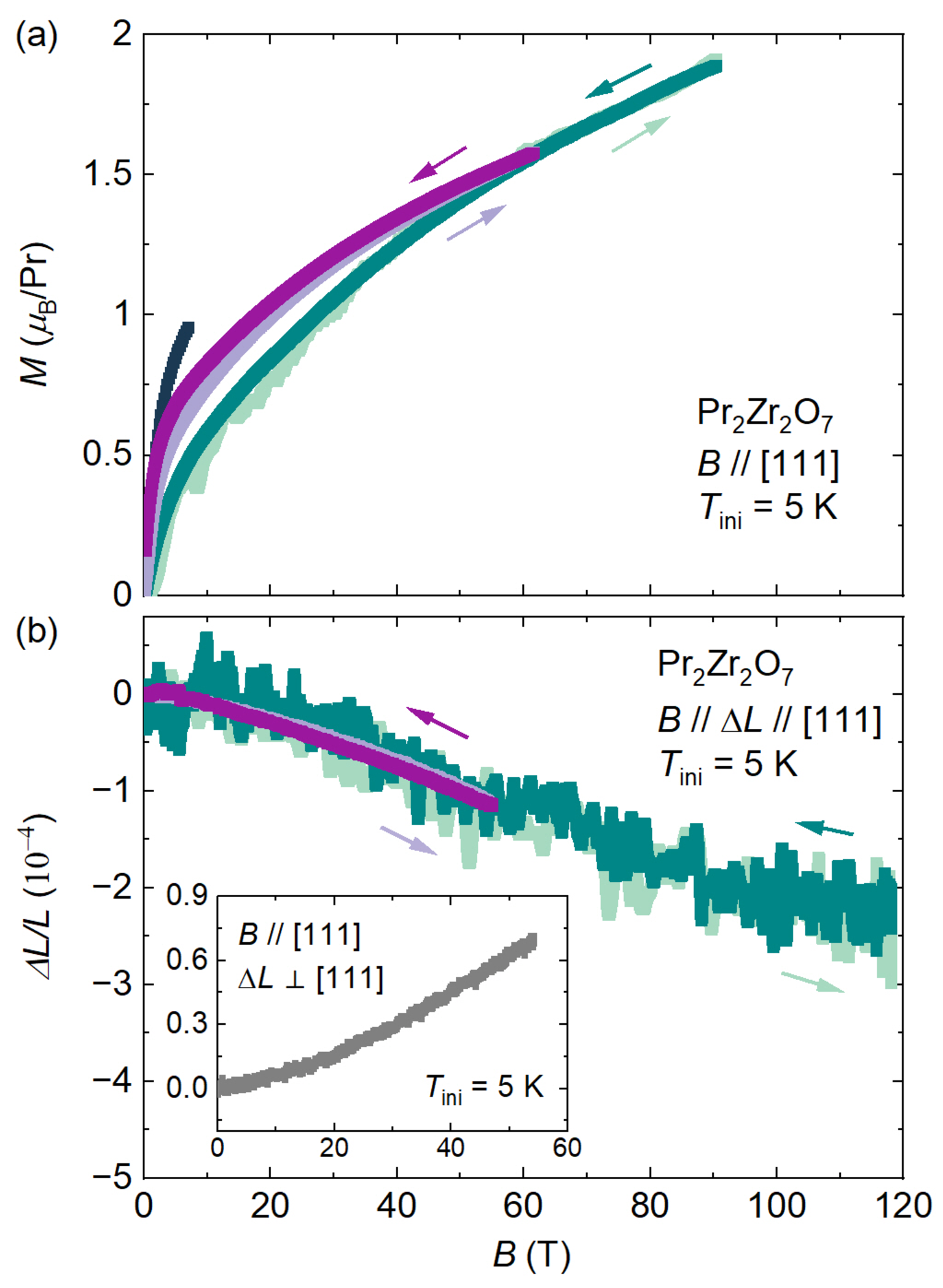}
\caption{Magnetic-field dependence of (a) magnetization $M$ and (b) longitudinal magnetostriction $\Delta L/L \parallel [111]$ in \PZO{} for $B \parallel [111]$ in ultrahigh magnetic fields. Black symbols in (a) show the $M$--$B$ data measured at $T = 4.2$~K using the MPMS, purple symbols show the data measured at $T_{\rm ini} = 4.2$~K using the nondestructive pulsed magnet, and green symbols show the data measured at $T_{\rm ini} = 5$~K using the HSTC system. The inset of (b) shows the transverse magnetostriction data measured at $T_{\rm ini} = 4.2$~K using the nondestructive pulsed magnet.}
\label{fig6}
\end{figure}

\vspace{-0.4cm}
\subsection{\label{Sec3F}Prevalence of CF striction in spin-ice compounds}

We have revealed the CF striction in \HTO{} through both experimental and theoretical approaches.
Applying a magnetic field along the [111] direction leads to the hybridization of the lowest and first-excited CF levels around $B_{\rm cf}$, which is associated with negative longitudinal magnetostriction and positive transverse magnetostriction.
To explore whether the CF striction is a common feature of spin-ice systems, we investigated high-field magnetostriction in another spin-ice compound, \PZO{}.
The introduction of rare-earth ions with smaller magnetic moments, such as Pr ($\sim$3~$\mu_{\rm B}$), can modify the spin-ice state, making it less Ising-like and more susceptible to quantum effects.
In particular, Pr$_{2}$(Zr, Sn, Hf, Ir)$_{2}$O$_{7}$ are proposed as quantum spin ices \cite{NTang_2023, HDZhou_2008, Machida_2010, KKimura_2013, Petit_2016, Sibille_2018}.
The lowest CF level of \PZO{} consists of 90--96\% of $|J=\pm 4 \rangle$ \cite{KKimura_2013, Petit_2016} (smaller compared to 98\% of $|J=\pm 8 \rangle$ in \HTO{}), indicating the presence of substantial transverse fluctuations.

Figures~\ref{fig6}(a) and \ref{fig6}(b) display the $M$--$B$ and $\Delta L/L$--$B$ curves for \PZO{} at $T_{\rm ini} = 5$~K (or 4.2~K), respectively.
No hysteresis is observed, consistent with the reported fast spin dynamics on the picosecond timescale \cite{KKimura_2013}.
Unlike \HTO{}, the magnetization shows a monotonic increase without any plateau-like features, suggesting that \PZO{} does not exhibit a well-defined Ising limit state, and the low-energy CF levels gradually hybridize as the magnetic field increases.

Notably, the longitudinal magnetostriction in \PZO{} decreases monotonically, while the transverse magnetostriction increases.
This anisotropic lattice deformation is similar to that observed in \HTO{} (Fig.~\ref{fig4}), suggesting that the similar CF environments in \HTO{} and \PZO{} lead to qualitatively identical CF striction effects.
We note that the sign of the magnetostriction in the low-field regime for \PZO{} is opposite to that for \HTO{}.
This difference is likely due to the opposite signs of $dJ/dR$ for \HTO{} ($dJ/dR > 0$) and \PZO{} ($dJ/dR < 0$).

\vspace{-0.2cm}
\section{Conclusion}

We have thoroughly investigated the static and dynamic magnetoelastic properties in the spin-ice compound Ho$_{2}$Ti$_{2}$O$_{7}$ through magnetostriction measurements under ultrahigh magnetic fields.
Our observations reveal a large anisotropic magnetostriction occurring at the CF level crossing around $B_{\rm cf} \sim 65$~T, where the longitudinal magnetostriction amounts to $\Delta L/L \sim -5 \times 10^{-4}$. 
Complementary MCE measurements reveal a dramatic sample-heating effect exceeding 20~K in pulsed high-field experiments, and also highlight the onset of CF level hybridization above 30~T, which would correlate with the turnaround behavior of the magnetostriction curve. 
By varying the magnetic field sweep rates, we distinguish between the fast dynamics of the CF-phonon coupling and the slower dynamics of spin-spin correlations, associated with monopole creation and annihilation.

To model the CF striction, we developed a comprehensive Hamiltonian based on a point-charge model that includes CF-phonon coupling.
Our numerical calculations based on a mean-field approach successfully reproduce the experimental magnetostriction data across the entire field range, showing a dominant contribution from CF striction $\Delta L/L_{\rm cf}$ and a much smaller contribution from exchange striction $\Delta L/L_{\rm ex}$. This study paves the way for understanding the intricate role of the CF-phonon coupling in the crystal structure of a broad range of $4f$-based magnets.

\vspace{-0.4cm}
\section*{ACKNOWLEDGMENTS}
This work was partly supported by the Deutsche Forschungsgemeinschaft (DFG, German Research Foundation) via TRR 360 (project No.~492547816), JSPS KAKENHI Grants-In-Aid for Scientific Research (Grants No.~20J10988 and No.~24H01633), JST-ASPIRE Program (JPMJAP2317), and JST-MIRAI Program (JPMJMI20A1). N.T. was supported by the Alexander von Humboldt foundation. M.G. was supported by the JSPS through a Grant-in-Aid for JSPS Fellows. We gratefully acknowledge the enduring support of T. Sakakibara and the stimulating discussions with R. Moessner, R. Klingeler, M. Gingras, T. Fennell, and M. Udagawa.

\appendix

\vspace{-0.4cm}
\section{\label{AppendixA}Waveform of pulsed high magnetic fields}

\begin{figure}[t]
\centering
\includegraphics[width=0.7\linewidth]{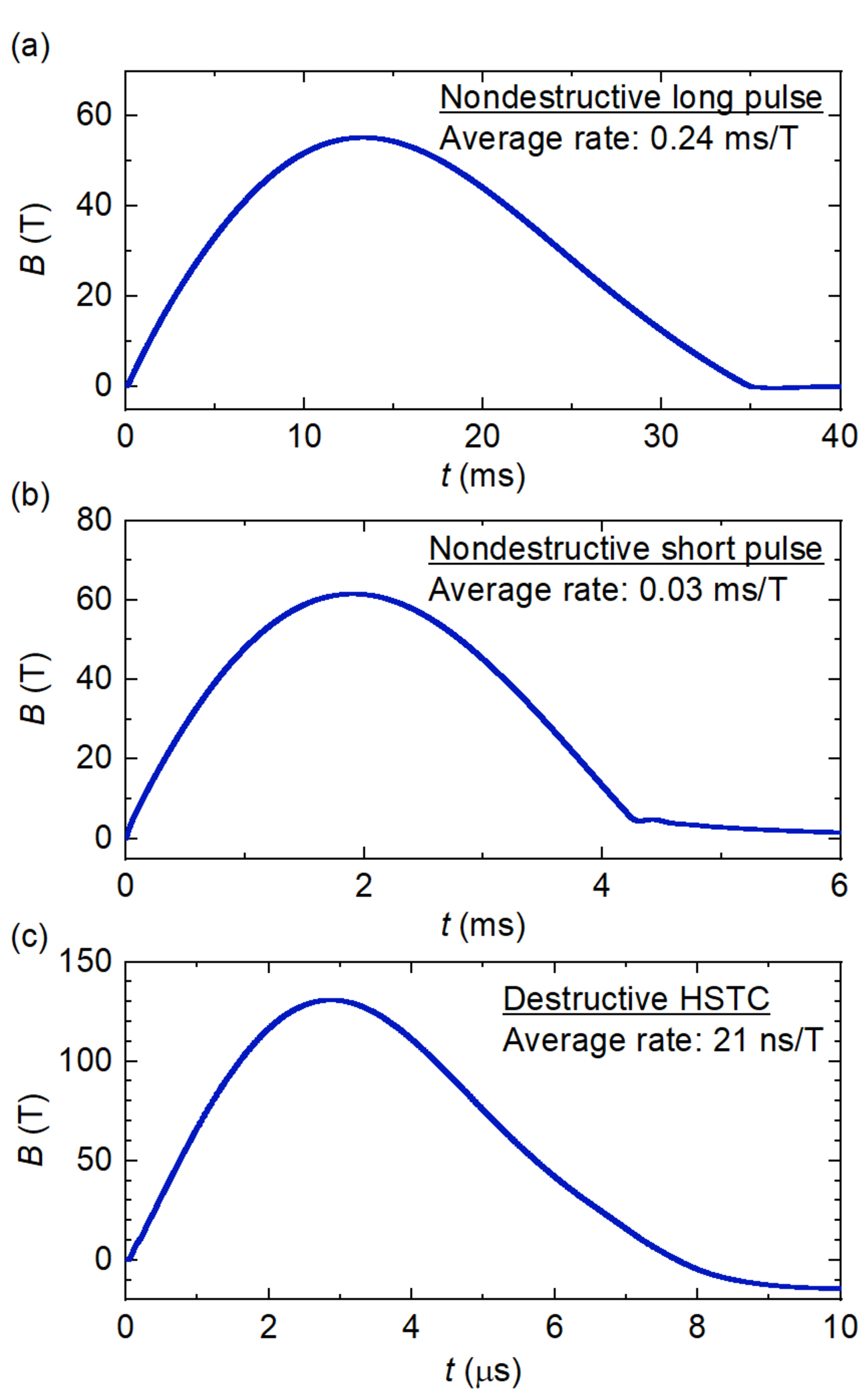}
\caption{Typical waveform of the generated pulsed magnetic field in (a) the nondestructive long pulsed magnet, (b) the nondestructive short pulsed magnet, and (c) the horizontal single-turn-coil (HSTC) system. The averaged field-sweep rates for the field-up sweep are 240~$\mu$s/T, 30~$\mu$s/T, and 21~ns/T, respectively.}
\label{fig7}
\end{figure}

Figure~\ref{fig7} shows the typical magnetic-field profiles generated by three types of pulsed magnets used in this study.

\vspace{-0.4cm}
\section{\label{AppendixB}Phonon and CF-phonon coupling terms}

\setcounter{equation}{0}\renewcommand\theequation{B\arabic{equation}}

In this section, we provide the formulas for the phonon term $\hat{H}_{\rm ph}$ and the CF-phonon coupling term $\hat{H}_{\rm cfph}$ in the Hamiltonian of Eq.~\eqref{eqn_H_total}.
$\hat{H}_{\rm ph}$ is written as follows:
\begin{equation}
\begin{aligned}
\label{eqn_H_ph}
\hat{H}_{\rm ph} =& \sum_i \frac{\bm{p}_i^2}{2m_i}
+\frac{1}{2}\sum_{ij} \frac{k_{ij}}{2|R_{ij}|^2}(\bm U_{j}\bm R_{ij}-\bm U_{i}\bm R_{ij})^2.\\
\end{aligned}
\end{equation}
Here, $\bm{p}_{i}$ and $m_{i}$ denote the momentum and mass, respectively, of the atom at position $i$.
The displacement vector of the atom at position $i$ is represented by $\bm U_i$.
The difference between the undisplaced lattice positions is given by $\bm R_{ij} \equiv \bm R_j-\bm R_i$, where $\bm R_j$ ($\bm R_i$) represents the position of atom at site $j$ ($i$).
$k_{ij}$ is the longitudinal atomic spring constants, which is determined based on the Born-von Karman model, assuming simple distance-dependence (see Appendix~\ref{AppendixC} for details).
The spring constants are then used to define the phonon-strain coupling constant $G_{\rm mix}$ as follows:
\begin{equation}
\begin{aligned}
\label{G_mix}
G_{\rm mix}^{\alpha \beta \gamma} (i)=&-a_0\sum_i \frac{k_{ij}}{|\bm R_{ij}|^2}\bm R_{ij}^{\alpha} \bm R_{ij}^{\beta} \bm R_{ij}^{\gamma},\\
\end{aligned}
\end{equation}
where $a_0$ is the Bohr radius, and the indices $\alpha, \beta, \gamma = x, y, z$ refer to an Euclidean coordinate system parallel to the crystal axes, i.e., $x \parallel a$, $y \parallel b$ and $z \parallel c$. To be noted, in Eq.~(\ref{eqn_strain_selfconsistent}), $\alpha$ and $\beta$ are used in Voigt notation ($\alpha, \beta =1, ..., 6$), and are defined differently than in this context. 

The CF-phonon coupling is expressed by
\begin{equation}
\label{eqn_H_cfph}
\hat{H}_{\rm cfph} =  \sum_{lm}\sum_{ij(i<j)}\nabla_{\bm U_i}B_{lm}(j){\bm U}_iO_{lm}(\hat{\bm J}_j),\\
\end{equation}
where the index $i$ denotes all atomic positions, $B_{lm}$ denotes the CF parameters, and $O_{lm}$ represents Stevens Operator equivalents.
The CF-phonon coupling constant $G_{\rm cfph}$ can be derived from Eq.~(\ref{eqn_H_cfph}) by taking the displacement derivative of the CF parameters as
\begin{equation}
\begin{aligned}
\label{eqn_G_cfph}
G_{\rm cfph}^{\alpha \beta lm} (j)= -\frac{1}{2}\sum_i (R_i^\beta\frac{\partial B_{lm}(j)}{\partial \bm{U}_i^\alpha}
+R_i^\alpha\frac{\partial B_{lm}(j)}{\partial \bm{U}_i^\beta}).
\end{aligned}
\end{equation}
Again, in Eq.~(\ref{eqn_strain_selfconsistent}), $\alpha$ and $\beta$ are used in Voigt notation ($\alpha, \beta =1, ..., 6$), and are defined differently than in this context. Importantly, $G{\rm _{cfph}}$ affects the magnitude of $\Delta L/L$.
We scaled $G{\rm _{cfph}}$ obtained from McPhase by a factor of 0.16 to match the calculated $\Delta L/L$ with the experimental one.
Although the elastic constants $c^{\alpha \beta}$ also affect the magnitude of $\Delta L/L$, their values were fixed based on prior experimental data, as detailed in Appendix~\ref{AppendixC}.

\begin{figure}[b]
\centering
\includegraphics[width=0.7\linewidth]{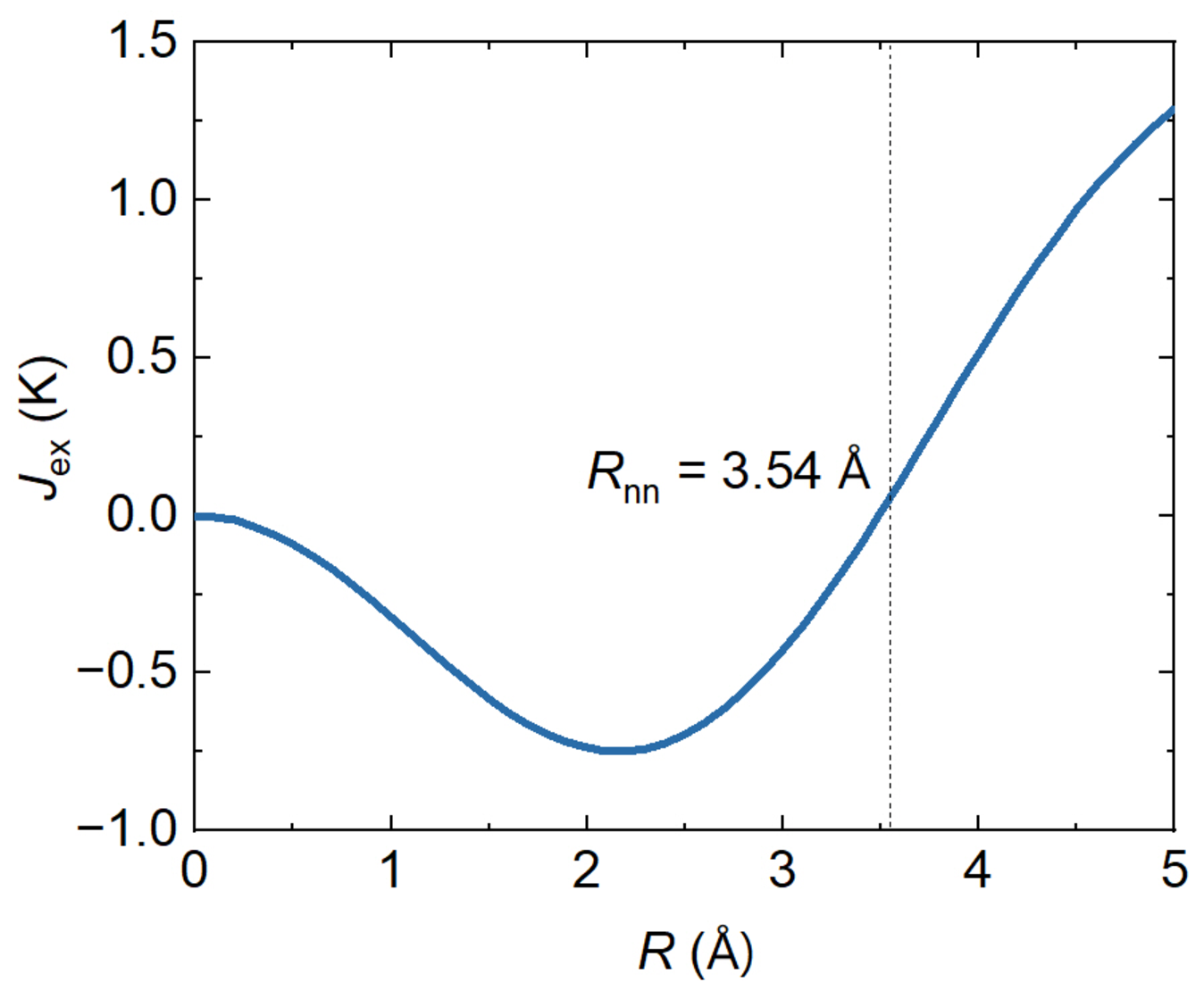}
\caption{Distance dependence of the nearest-neighbor exchange interaction $J_{\rm ex}$, as described by the Bethe-Slater curve. The nearest-neighbour distance is $R_{\rm nn}=3.54$~$\AA$ \cite{Baroudi_2015}.}
\label{fig8}
\end{figure}

In order to obtain the positional derivative of the exchange interaction $J_{\rm ex}$, the Bethe-Slater curve is employed, as shown in Fig.~\ref{fig8}: 
\begin{equation}
\label{eqn_JR}
J_{\rm ex}(R)=A[-(R/D)^2+(R/D)^4]\exp^{-(R/D)^2},
\end{equation}
where $A$ and $D$ are set to 0.4 and 3.49, respectively, resulting in $J_{\rm ex}=0.1$~K and $dJ_{\rm ex}/dr=1.03$~K/$\AA$ at the nearest-neighbor distance $R_{\rm nn}=3.54 $~$\AA$ \cite{Baroudi_2015}.
It is important to constrain $dJ_{\rm ex}/dr$ to a positive value for \HTO{} in order to explain the positive longitudinal magnetostriction in the low-field regime.

\vspace{-0.4cm}
\section{\label{AppendixC}Elastic constants}

\setcounter{equation}{0}\renewcommand\theequation{C\arabic{equation}}

\textsf{McPhase} computes the spring constants $k_{ij}$ in units of N/m based on the Born-von Karman model:
\begin{equation}
\label{eqn_C_long}
k_{ij} =k_{0} \times \exp(-gr^2), k_{0} = 800~{ \rm N/m}, g = 0.1 \AA ^{-2}
\end{equation}
where $i,j$ denotes spatial directions $x,y,z$, $k_{0}$ and $g$ are adjustable fitting coefficients used to obain the appropriate ealstic constants, and $r$ is bond length in the unit of $\AA$. Springs for neighbours up to a distance of 7.5~$\AA$ are taken into account. Based on these spring constants, the elastic constant tensor $c^{\alpha \beta}$ can be derived.
The resulting tensor closely matches those obtained from sound velocity experiments \cite{YNakanishi_2011} and successfully reproduces both the slope of low-energy acoustic phonons as a function of wave vectors and the highest-energy optical phonon mode, which reaches 105~meV in DFT calculations \cite{MRuminyDFT_2016}.
Equation~\eqref{eqn_C_long} produces a table of longitudinal spring constants, which are then converted to an elastic tensor with three independent elements in the units of eV/primitive unit cell: $c_{11}=585.656$, $c_{12}=276.629$, and $c_{44}=276.629$, reflecting the highly symmetric cubic crystal structure.
These values are in good agreement with those obtained from ultrasound measurements \cite{YNakanishi_2011}, after performing a unit conversion of 1~meV$/$primitive unit cell volume = 0.000621~GPa, where the primitive unit cell volume is 257.835~$\AA^3$ \cite{Baroudi_2015}.

\end{document}